\documentclass[11pt]{article}
\usepackage{amsmath,color}
\usepackage{amssymb}
\usepackage{braket}
\usepackage{acro}

\def\dd{\mathrm d}
\def\dd{\mathrm d}
\DeclareAcronym{cft}{
    short = {CFT},
    long = {conformal field theory},
    tag = {a}
}

\DeclareAcronym{qt}{
    short = {QT},
    long = {quasi-topological},
    tag = {a}
}

\begin{document}

\begin{center}
    {\Large\bf Holographic Three-Point Functions from Higher Curvature Gravities in Arbitrary Dimensions}

    \bigskip

    Fei-Yu Chen\textsuperscript{1} and H.~L\"u\textsuperscript{1,2}

    \bigskip

    \textsuperscript{1}{\it Center for Joint Quantum Studies and Department of Physics,\\
    School of Science, Tianjin University, Tianjin 300350, China}

    \bigskip

    \textsuperscript{2}{\it Joint School of National University of Singapore and Tianjin University,\\
    International Campus of Tianjin University, Binhai New City, Fuzhou 350207, China}
\end{center}

\bigskip

\begin{abstract}

We calculate the holographic three-point function parameters $\mathcal A$, $\mathcal B$, $\mathcal C$ in general $d\geqslant 4$ dimensions from higher curvature gravities up to and including the quartic order. The result is valid both for massless and perturbative higher curvature gravities. It is known that in four dimensional CFT the $a$-charge is a linear combination of $\mathcal A$, $\mathcal B$, $\mathcal C$, our result reproduces this but also shows that a similar relation does not exist for general $d > 4$. We then compute the Weyl anomaly in $d = 6$ and found all the three $c$-charges are linear combinations of $\mathcal A$, $\mathcal B$, $\mathcal C$, which is consistent with that the $a$-charge is not. We also find the previously conjectured relation between $t_2$, $t_4$, $h''$ does not hold in general massless gravities, but holds for quasi-topological ones, and we obtain the missing coefficient.

\end{abstract}

\newpage

\section{Introduction}

In \ac{cft}, conformal invariance typically requires the correlation functions to have fairly rigid forms. For example, in flat spacetime the two-point function of the energy-momentum tensor $T_{ab}$ is completely determined by the parameter $C_T$ \cite{Osborn:1993cr, Erdmenger:1996yc}
\begin{equation}\label{eq:cft-t-2pt}
    \braket{0|T_{ab}(x)T_{cd}(y)|0} = C_T \frac{\mathcal I^{(0)}_{abcd}(x - y)}{|x - y|^{2d}},
\end{equation}
where $d$ is the spacetime dimension. Similarly, after imposing further constraints that arise from the conservation of energy, the three-point function of $T_{ab}$ is controlled by three parameters $\mathcal A$, $\mathcal B$, $\mathcal C$ \cite{Osborn:1993cr, Erdmenger:1996yc}, {\it i.e.}
\begin{equation}
    \braket{0|T_{ab}(x)T_{cd}(y)T_{ef}(z)|0} = \frac{\mathcal A\,\mathcal I^{(1)}_{abcdef} + \mathcal B\,\mathcal I^{(2)}_{abcdef} + \mathcal C\,\mathcal I^{(3)}_{abcdef}}{|x - y|^d |y - z|^d |z - x|^d}.
\end{equation}
Note that the $\mathcal I^{(0)}, \mathcal I^{(1)}, \mathcal I^{(2)}$ and $\mathcal I^{(3)}$ are tensorial structures whose explicit forms are inessential in our discussion here. These parameters are generally independent in $d \geqslant 4$, but for $d = 3$ and $d = 2$, the ${\cal I}^{(i)}$ tensors  become degenerate and the number of independent coefficients is two and one respectively.

In a curved spacetime background, \ac{cft} in even dimensions becomes anomalous in that the trace of $T_{ab}$  acquires a non-zero expectation value known as the trace anomaly \cite{Duff:1977ay}
\begin{equation}\label{eq:cft-trace-anomaly}
    (4\pi)^{d/2}\braket{T_a^a} = -a E^{(d)} + \sum_i c_i I_i^{(d)},
\end{equation}
where $E^{(d)}$ and $I_i^{(d)}$ are the Euler density and Weyl invariants in (even) $d$ dimensions respectively. The coefficients $a$ and $c_i$ are known as central charges. In general even dimensions, $C_T$ is a linear combination of the $c_i$'s \cite{Osborn:1993cr}; it is thus simply proportional to the only $c$ in $d=4$. Although the concept of conformal anomaly no longer applies in odd dimensions, the quantity $C_T$, a linear combination of $c_i$'s in even dimensions, survives. The $a$-charge can also be generalized to odd dimensions as the universal coefficient of the entanglement entropy across a spherical entangling surface, which coincides with $a$-charge in even dimensions \cite{Casini:2011kv}. Thus the $a$-charge, $C_T$, and three-point function parameters ($\mathcal A$, $\mathcal B$, $\mathcal C$) are important characteristics of \ac{cft} in general dimensions as they control the energy-momentum tensor correlators up to and including three points. These parameters are not independent. It was shown by Ward identity that $C_T$ is a linear combination of $\mathcal A$, $\mathcal B$, $\mathcal C$ \cite{Osborn:1993cr, Erdmenger:1996yc}, namely
\begin{equation}\label{eq:ct-from-abc}
    C_T = \frac{\pi^d}{\Gamma(d/2)}\frac{(d - 1)(d + 2)\mathcal A - 2\mathcal B - 4(d + 1)\mathcal C}{d(d + 2)}.
\end{equation}
In $d = 4$, a similar relation exists for $a$ \cite{Osborn:1993cr}
\begin{equation}\label{eq:a-charge-to-abc}
    a = \frac{\pi^6}{2880}(13\mathcal A - 2\mathcal B - 40\mathcal C)\,.
\end{equation}
It thus is tempting to expect that analogous relation of \eqref{eq:ct-from-abc} would exist also for the $a$-charge.

The AdS/CFT correspondence provides a new way to study the large $N$ limit of \ac{cft}s from the weakly-coupled bulk gravity theory in the anti-de sitter (AdS) background in $D=d+1$ dimensions \cite{Maldacena:1997re}. Einstein gravity extended with higher-order curvature invariants is insightful to study as they capture new features of the dual \ac{cft}. The large number of higher-derivative terms provide unlimited data that can not only reveal the CFT in the large $N$ limit, but also some universal properties of \ac{cft}.

There has been extensive studies of the holographic correlators in the pure gravity sector with higher curvature extensions, e.g. ref.\cite{Buchel:2009sk, Myers:2010jv,Bueno:2018xqc, Li:2019auk,Camanho:2013pda}. For general higher curvature gravity, the corresponding linearized equation of motion around AdS background contains at most four derivatives of the metric, the graviton spectrum thus contains two extra modes, the massive scalar mode and ghost-like spin-2 mode \cite{Bueno:2016ypa}. Unitarity of the dual \ac{cft} requires the decoupling of the ghost mode, while the decoupling of the scalar mode is required by the holographic $a$-theorem \cite{Li:2017txk}. Thus we usually require the decoupling of both modes, either exactly or perturbatively, resulting in massless gravity or perturbative gravity, respectively. Due to computational difficulties, three-point function parameters of higher curvature gravity are usually calculated indirectly {\it via} the energy flux parameters $t_2$ and $t_4$ \cite{Myers:2010jv, Hofman:2008ar}, together with $C_T$, they contain all the information about the parameters $\cal (A, B, C)$. Three-point function parameters for massless cubic gravity was calculated in ref.\cite{Li:2019auk} for $d = 3$ and $d = 4$, while the results in arbitrary dimensions or with higher-order curvature invariants are still absent in the literature.

Several features of $(a, C_T)$ and $\cal (A, B, C)$ have been known. It was shown that to distinguish $a$ and $c$ holographically in $d = 4$, one needs to introduce at least the Gauss-Bonnet density, in which case only two of the three three-point function parameters are independent; one needs to further consider cubic curvature invariants to obtain all the three independent parameters \cite{Camanho:2013pda}. In additional to these algebraic relations, holography also provides a hidden differential relation between the $c$ and $a$ charges for all massless higher curvature gravities \cite{Li:2018drw, Li:2021jfh}
\begin{equation}\label{eq:ct-a-hidden-rel}
    C_T = \frac{\Gamma(d + 2)}{\pi^{d}(d - 1)^2}L\frac{\partial a}{\partial L}\,,
\end{equation}
where $L$ is the effective radius of AdS. Furthermore, it was conjectured that a relation exists between $t_2$, $t_4$, $h''$ for massless gravity \cite{Bueno:2018yzo}
\begin{equation}\label{eq:t2-t4-hpp-relation}
    a_{(d)}t_2 + b_{(d)}t_4 = -\frac{h''(-L^{-2})}{L^2 h'(-L^{-2})}\,,
\end{equation}
where $h(\lambda)$ is the AdS vacuum equation of motion, to be defined later. This relation was verified for Lovelock gravity in arbitrary dimensions and quasi-topological gravity in $d = 4$. In these verifications, only $a_{(d)}$ and $b_{(4)}$ were determined, since for Lovelock gravity we have $t_4 = 0$. It is thus interesting to check for more general curvature invariants in diverse dimensions and, if possible, to determine the value of $b_{(d)}$.

The main purpose of this work is to calculate holographically the parameters ($\mathcal A$, $\mathcal B$, $\mathcal C$) of the three-point function from the general higher curvature invariants up to and including the quartic order in arbitrary $D=d+1$ dimensions with $d \geqslant 4$. We employ an indirect method by considering the one-point function of an energy flux operator parametrized by $t_2$ and $t_4$, which are directly related to the three-point function parameters. We consider both the finite higher-derivative coupling case and perturbative case. For the former, we need to impose the massless conditions. For the latter the number of non-trivial terms can be reduced by some appropriate order-by-order field redefinitions of the metric.

From our results, we can make some interesting statements:
\begin{enumerate}
    \item After setting $d = 3$ for the general $d$ results, the value of $t_4$ coincides with that obtained in ref.\cite{Li:2019auk}, even though our method should only be valid for $d \geqslant 4$.

    \item The $a$-charge and ($\mathcal A$, $\mathcal B$, $\mathcal C$) turn out to be in general linearly independent for $d \geqslant 5$. In other words, there is no generalization of \eqref{eq:a-charge-to-abc} beyond $d=4$.

    \item Instead, all the three $c$-charges in $d = 6$ are linear combinations of ($\mathcal A$, $\mathcal B$, $\mathcal C$).

    \item The conjectured relation \eqref{eq:t2-t4-hpp-relation} does not hold for general massless gravity, but may hold for quasi-topological gravities. We verify this up to and including the quartic order.
\end{enumerate}
The three-point function, and hence $\cal (A, B, C)$, contains three different possible structures, which can be enumerated holographically by Einstein, quadratic, and cubic curvature terms \cite{Camanho:2013pda,Afkhami-Jeddi:2016ntf,Meltzer:2017rtf}. It was claimed by ref.\cite{Li:2019auk} without proof that higher order curvature polynomials will not provide further information. In fact, this can be easily proved, and we present it in appendix \ref{sec:ap:general-3pt}. Nevertheless, the quartic-order calculation provides a useful consistent check.

The paper is organized as follows. In section \ref{sec:higher-curvature-grav-rev} we review higher curvature gravity and list all the Reimann curvature polynomial invariants up to and including the quartic order. In section \ref{sec:efparams-calc} we first briefly review the method we use to calculate the energy flux parameters, following ref.\cite{Myers:2010jv}, then we present our results and cross check them with the known special cases. Further discussions on our results are given in section \ref{sec:discussions}, where we also calculate the $a$-charge in general dimensions and the three $c$-charges in $d = 6$. We conclude this paper and make further comments in section \ref{sec:conclusion}. Some lengthy expressions and digressions are given in five appendix sections.

\section{A brief review on higher curvature gravities}
\label{sec:higher-curvature-grav-rev}

We consider Einstein gravity extended with higher-order curvature invariant polynomials in $D=d + 1$ dimensions, up to and including the quartic order. The general form of action is
\begin{equation}
    S = \int\dd^{d + 1}x \sqrt{|g|}\, \mathcal L(R_{abcd}, g_{ab}) = \int\dd^{d + 1}x\sqrt{|g|}\left[R + \frac{d(d - 1)}{L_0^2} + \sum_{i, j} e_{i, j}\mathcal R^{(i)}_j\right],\label{genaction}
\end{equation}
where $\mathcal R^{(i)}_j$ is the $j$'th Riemann scalar polynomial of order $i$, coefficients $e_{i, j}$ are the coupling constants, and $L_0$ is the bare AdS radius. All the Riemann curvature polynomial invariants studied in this paper are given explicitly below
\begin{align}
    \mathcal R^{(2)} & = \{R^2, \, R_{ab}R^{ab}, \, R_{abcd}R^{abcd}\}\,,\\
    \mathcal R^{(3)} & = \{R^3, \, R_{ab}R^{ab}R, \, R_a{}^b R_b{}^c R_c{}^a, \, R_{abcd}R^{ac}R^{bd}, \, RR_{abcd}R^{abcd}, \, R^{ab}R_a{}^{cde}R_{bcde}\,,\nonumber\\
    & \qquad R_{ab}{}^{cd}R_{cd}{}^{ef}R_{ef}{}^{ab}, \, R_a{}^e{}_c{}^fR^{abcd}R_{bedf}\},\\
    \mathcal R^{(4)} & = \{R^4, R_{ab}R^{ab}R^2, R_a{}^c R^{ab}R_{bc}R, (R_{ab}R^{ab})^2, R_a{}^c R^{ab}R_b{}^d R_{cd}, R R^{ab} R^{cd} R_{acbd}\,,\nonumber\\
    & \qquad R^{ab}R_c{}^eR^{cd}R_{adbe}, R^2 R_{abcd}R^{abcd}, RR^{ab}R_{cdeb}R^{cde}{}_a, R_{ab}R^{ab} R_{cdef}R^{cdef},\nonumber\\
    & \qquad R_a{}^c R^{ab} R_{defc} R^{def}{}_b, R^{ab}R^{cd}R_{efbd}R^{ef}{}_{ac}, R^{ab}R^{cd}R_{ecfd}R^e{}_a{}^f{}_b, R^{ab}R^{cd}R_{ebfd}R^e{}_a{}^f{}_c\,,\nonumber\\
    & \qquad R R_{ab}{}^{ef} R^{abcd}R_{cdef}, R R_a{}^e{}_c{}^f R^{abcd} R_{bedf}, R^{ab}R_a{}^c{}_b{}^d R_{efgd}R^{efg}{}_c\,, \nonumber\\
    & \qquad R^{ab}R_{cd}{}^g{}_a R^{cdef}R_{efgb}, R^{ab}R_c{}^g{}_{ea}R^{cdef}R_{dgfb}, (R_{abcd}R^{abcd})^2\,, \nonumber\\
    & \qquad R_{abc}{}^e R^{abcd}R_{fghe}R^{fgh}{}_d, R_{ab}{}^{ef}R^{abcd}R_{cdgh}R_{ef}{}^{gh}, R_{ab}{}^{ef}R^{abcd}R_{ce}{}^{gh}R_{dfgh}\,,\nonumber\\
    & \qquad R_{ab}{}^{ef}R^{abcd}R_c{}^g{}_e{}^h R_{dgfh}, R_a{}^e{}_c{}^f R^{abcd}R_{bgdh}R_e{}^g{}_f{}^h, R_a{}^e{}_c{}^f R^{abcd}R_e{}^g{}_b{}^h R_{fgdh}\}.
\end{align}
In other words, there are 3, 8, 26 terms for the quadratic, cubic and quartic orders respectively. The indices $(i,j)$ of the coefficients $e_{i,j}$ in \eqref{genaction} are labeled based on the above order.

Before proceeding, we shall briefly review some general properties of higher curvature gravity. The equation of motion is given by $\mathcal E^{ab}=0$, with
\begin{equation}\label{eq:general-eom}
    \mathcal E^{ab} = -\frac{1}{\sqrt{|g|}}\frac{\delta S}{\delta g_{ab}} = P^a{}_{cde}R^{bcde} - \frac12 g^{ab}\mathcal L - 2\nabla_c\nabla_d P^{acdb}, \qquad P^{abcd} = \frac{\partial\mathcal L}{\partial R_{abcd}}\,.
\end{equation}
The equation admits the maximally-symmetric solution $\bar g_{ab}$ whose Riemann tensor is given by $R_{abcd} = 2 \lambda \, \bar g_{a[c}\bar g_{d]b}$, where $\lambda$ is to be determined. When evaluated on such a background, the tensor $P^{abcd}$ takes the following simple form \cite{Bueno:2016xff}
\begin{equation}\label{eq:l-pd-r-form}
    \left.P^{abcd}\right|_{\bar g_{ab}} = 2 k_1 \bar g^{a[c}\bar g^{d]b}.
\end{equation}
Substituting into \eqref{eq:general-eom} some useful identities emerge \cite{Bueno:2016xff, Bueno:2018yzo}
\begin{align}
    \left.\mathcal E^{ab}\right|_{\bar g_{ab}} & = \left(2d \lambda k_1 - \frac12 \left.\mathcal L\right|_{\bar g_{ab}}\right)\bar g^{ab} = 0 \implies \mathcal L(\lambda) \equiv \left.\mathcal L\right|_{\bar g_{ab}} = 4d\lambda k_1,\\
    \mathcal L'(\lambda) & = \left(P^{abcd} \frac{\dd R_{abcd}}{\dd \lambda}\right)_{\bar g_{ab}} = 2d(d + 1)k_1,\\
    h(\lambda) & = \mathcal L(\lambda) - \frac{2\lambda}{d + 1}\mathcal L'(\lambda) = 0\,. \label{eq:ads-vacuum-eqn}
\end{align}
The last equation determines the on-shell value of $\lambda$. Note that when taking a derivative with respect to $\lambda$, all the parameters in the theory including $L_0$ should be treated as being independent of $\lambda$.

The linearized theory is governed by the tensor
\begin{equation}\label{eq:lpdr3-c-def}
    C^{abcdefgh} = \frac{\partial^2\mathcal L}{\partial R_{abcd}\partial R_{efgh}}.
\end{equation}
Evaluating on the maximally-symmetric background, it takes the form \cite{Bueno:2016xff}
\begin{align}
    \left.C^{abcdefgh}\right|_{\bar g_{ab}} & = k_2 \left(\bar g^{[a|e|}\bar g^{b]f}\bar g^{[c|g|}\bar g^{d]h} + \bar g^{[a|g|}\bar g^{b]h}\bar g^{[c|e|}\bar g^{d]f}\right)\nonumber\\
    & \qquad + 4k_3 \, \bar g^{a[c}\bar g^{|b|d]}\bar g^{e[g}\bar g^{|f|h]} + 4k_4 \delta^{[a}_{(p}\bar g^{b][c}\delta^{d]}_{q)}\bar g^{p[e}\bar g^{f][g}\bar g^{|q|h]}\label{eq:l-pd-r2-form}.
\end{align}
To be precise, the above tensor structure does not satisfy the cyclic Bianchi identity inherited from the Riemann tensor.  We should impose this identity and redefine the tensor as follows
\begin{equation}
C^{abcdefgh}\quad \rightarrow \quad C^{abcdefgh}=C^{abcdefgh}- C^{a[bcd]e[fgh]}\,.
\end{equation}
The linearized theory of any higher curvature theory is completely described by the coefficients $k_i$, $i=1,2,3,4$. For a Lagrangian that is a polynomial of curvature invariants, these four coefficients are linear functions of coupling constants of the polynomial invariants. For a specific Lagrangian, the coefficients $k_i$ can be obtained efficiently with the method proposed in ref.\cite{Bueno:2016xff}. The effective Newton constant $\kappa_{\mathrm{eff}}$ and the masses of the scalar mode $m_s$ and ghost-like spin-2 mode $m_g$ are given below \cite{Bueno:2016xff}
\begin{align}
    \kappa_{\mathrm{eff}}^{-1} & = 4k_1 - 8l(d - 2)k_2\,,\\
    m_g^2 & = \frac{-k_1 + 2(d - 2)lk_2}{2k_2 + k_4}\,,\qquad
    m_s^2 = \frac{(d - 1)k_1 - 4l[k_2 + d(d + 1)k_3 + d k_4]}{2k_2 + 4d k_3 + (d+ 1)k_4}\,.
\end{align}
The quantity $1/\kappa_{\mathrm{eff}}$ appears as a coefficient of the linearized equation of motion of the massless graviton after the massive modes are decoupled, namely
\begin{equation}
    \mathcal E^{ab}_L \sim \frac{1}{\kappa_{\mathrm{eff}}}(\Box - 2\lambda)h^{ab}.
\end{equation}
The central charge $C_T$ of the dual CFT is related to $\kappa_{\mathrm{eff}}$ as
\begin{equation}\label{eq:ct-from-kappaeff}
    C_T = \frac{\Gamma(d + 2)}{\pi^{d/2}(d - 1)\Gamma(d/2)}\frac{L^{d - 1}}{\kappa_{\mathrm{eff}}}\,.
\end{equation}
To decouple both the massive modes, we require
\begin{equation}
    2k_2 + k_4 = 2k_2 + 4d k_3 + (d+ 1)k_4 = 0\,.
\end{equation}
At the quadratic order, we have $k_1=\lambda d(d+1)e_{2, 1} + \lambda d e_{2, 2} + 2\lambda e_{2, 3}$, $k_2 = e_{2, 3}$, $k_3 =  e_{2, 1}/2$ and $ k_4 = e_{2, 2}/2$, the above conditions yield precisely the Gauss-Bonnet density. At the cubic order, we have the following constraints on the coupling constants \cite{Li:2017txk}
\begin{align}
    d (d+1) e_{3,2}+3 d e_{3,3}+(2 d-1) e_{3,4}+4 d (d+1) e_{3,5}+4 (d+1) e_{3,6}+24 e_{3,7}-3 e_{3,8} & = 0,\nonumber\\
    12 d (d+1) e_{3,1}+d (d+9) e_{3,2}+3 d e_{3,3}+(2 d+3) e_{3,4}+16 e_{3,5}+4 e_{3,6}+3 e_{3,8} & = 0.\label{cubicmassless}
\end{align}
For the quartic order the constraints become too lengthy and we record them in appendix \ref{sec:ap:quartic-massless-cond}.

An alternative approach to higher-order gravity is to treat it as an effective theory of quantum corrections to Einstein gravity. In this approach, the massive modes are decoupled perturbatively since at the zeroth order the coefficient $k_1$ is nonzero while $k_{2, 3, 4}$ are first order, making the kinetic terms of the massive modes first order. This applies to effective field theory where the first massive state appear at some high energy cutoff scale $M$ \cite{Caron-Huot:2021enk}. In this approach, one can perform field redefinitions of the metric order-by-order by appropriate higher curvature terms, so as to  eliminate some Ricci tensor and scalar terms in the Lagrangian. This gives the following residual sets of Riemann polynomials:
\begin{align}
    \mathcal R'^{(2)} & = \{R_{abcd}R^{abcd}\},\\
    \mathcal R'^{(3)} & = \{R_{ab}{}^{cd}R_{cd}{}^{ef}R_{ef}{}^{ab}, \, R_a{}^e{}_c{}^fR^{abcd}R_{bedf}\},\\
    \mathcal R'^{(4)} & = \{(R_{abcd}R^{abcd})^2, R_{abc}{}^e R^{abcd}R_{fghe}R^{fgh}{}_d, R_{ab}{}^{ef}R^{abcd}R_{cdgh}R_{ef}{}^{gh},\nonumber\\
    & \qquad R_{ab}{}^{ef}R^{abcd}R_{ce}{}^{gh}R_{dfgh}, R_{ab}{}^{ef}R^{abcd}R_c{}^g{}_e{}^h R_{dgfh}, R_a{}^e{}_c{}^f R^{abcd}R_{bgdh}R_e{}^g{}_f{}^h,\nonumber\\
    & \qquad R_a{}^e{}_c{}^f R^{abcd}R_e{}^g{}_b{}^h R_{fgdh}\}\,.
\end{align}
The coupling constants of these terms are invariant under the field redefinition.

In many of our calculations in this paper, we find that it is not necessary to impose the massless conditions explicitly. Therefore many of our results are valid for both approaches to higher-derivative gravities.

\section{Holographic calculation of energy flux parameters}
\label{sec:efparams-calc}

We now turn to the main subject of this work. We shall determine holographically the three-point function parameters ($\mathcal A$, $\mathcal B$, $\mathcal C$) of the energy-momentum tensor in the dual \ac{cft}. To calculate the three-point function directly from higher curvature gravities, one needs to perturb the metric to the third order in the Lagrangian, which is quite challenging even for Einstein gravity \cite{Arutyunov:1999nw}. We therefore employ an alternative way to determine these parameters.

We follow ref.\cite{Myers:2010jv,Buchel:2009sk} by considering a specific frame and polarization in which the three-point function describes a hypothetical conformal collider experiment proposed in ref.\cite{Hofman:2008ar}. In this experiment one first creates a localized excitation with the operator $\mathcal O \sim \varepsilon_{ij} T^{ij}$ where $\varepsilon_{ij}$ is the polarization tensor, then measures the energy flux at the null infinity of the direction indicated by the unit vector $\vec n$. The energy-flux operator $\mathcal E(\vec n)$ is
\begin{equation}
    \mathcal E(\vec n) = \lim_{r \to +\infty} \int_{-\infty}^{+\infty}\dd t \, T^t{}_i(t, r\vec n)n_i\,.
\end{equation}
Its expectation value takes the form
\begin{equation}\label{eq:energy-flux-op}
    \braket{\mathcal E(\vec n)} = \frac{\braket{0|\mathcal O^\dagger\mathcal E(\vec n)\mathcal O|0}}{\braket{0|\mathcal O^\dagger\mathcal O|0}}\,.
\end{equation}
For $d\geqslant 3$, by $\mathrm O(d - 1)$ invariance the most general form of the energy flux can be determined by two parameters $t_2$ and $t_4$ \cite{Hofman:2008ar}
\begin{equation}\label{eq:energy-flux-general-form}
    \braket{\mathcal E(\vec n)} = \frac{E}{\Omega_{d - 1}}\left[1 + t_2\left(\frac{\varepsilon_{ij}^\ast\varepsilon_{il}n^jn^l}{\varepsilon_{ij}^\ast\varepsilon_{ij}} - \frac{1}{d - 1}\right) + t_4\left(\frac{\left|\varepsilon_{ij}n^in^j\right|^2}{\varepsilon_{ij}^\ast\varepsilon_{ij}} - \frac{2}{d^2 - 1}\right)\right].
\end{equation}
Note that for $d=3$, the coefficient of $t_2$ vanishes identically and we are left with only the $t_4$ term. To isolate the contribution from null infinity, it is convenient to define a new set of coordinate
\begin{equation}\label{eq:bdr-y-to-x}
    y^+ = -\frac{L^2}{x^+}, \qquad y^- = x^- - \frac{x^{\bar i}x_{\bar i}}{x^+}, \qquad y^{\bar i} = L\frac{x^{\bar i}}{x^+},
\end{equation}
where $x^\pm = t \pm x^{d - 1}$, $L$ is some energy scale, to be chosen as the bulk AdS radius, and $\bar i$ denotes the index of the $(d - 2)$ dimensional subspace, {\it i.e.}, $1 \leqslant \bar i \leqslant d - 2$. This is in fact a conformal transformation with conformal factor $(y^+ / L)^2$, after which the energy-flux operator becomes
\begin{equation}\label{eq:energy-flux-op-in-y}
    \mathcal E(\vec n) = L^2\Omega^{d - 1}\int_{-\infty}^{+\infty}\dd y^- T_{--}(y^+ = 0, y^-, y^{\bar i} = y'^{\bar i})\,,
\end{equation}
where
\begin{equation}
    \Omega = \frac{2}{1 + n^{d - 1}}\,, \qquad y'^{\bar i} = \frac{L n^{\bar i}}{1 + n^{d - 1}}\,.
\end{equation}
On the other hand, the excitation operator $\mathcal O$ takes the form
\begin{equation}
    \mathcal O = \int\dd^{d}x \, e^{-iEt}\varepsilon_{ij}T^{ij}\psi(x/\sigma)\,,
\end{equation}
where $\psi(x)$ is some distribution that's localized at $x = 0$, and $E\sigma \gg 1$ is assumed so that the operator is localized. Thus one can see that the numerator of \eqref{eq:energy-flux-op} is indeed the three-point function with indices contracted with specific polarizations, thus one can relate ($t_2$, $t_4$) to ($\mathcal A$, $\mathcal B$, $\mathcal C$) \cite{Buchel:2009sk}
\begin{align}
    t_2 & = \frac{(2 (d+1)) \left(\mathcal A (d-2) (d+1) (d+2)+3 \mathcal B d^2-4 \mathcal C (2 d+1) d\right)}{d (\mathcal A (d-1) (d+2)-2 \mathcal B-4 \mathcal C (d+1))},\label{eq:t2-from-abc}\\
    t_4 & = -\frac{(d+1) \left(\mathcal A \left(2 d^2-3 d-3\right) (d+2)+2 \mathcal B (d+2) d^2-4 \mathcal C (d+1) (d+2) d\right)}{d (\mathcal A (d-1) (d+2)-2 \mathcal B-4 \mathcal C (d+1))}.\label{eq:t4-from-abc}
\end{align}
Using the above and \eqref{eq:ct-from-abc} one can solve the parameters  ($\mathcal A$, $\mathcal B$, $\mathcal C$) in terms of ($t_2$, $t_4$, $C_T$), thus the problem is converted to the calculation of $(t_2,t_4,C_T)$ quantities.

To calculate the energy flux parameters holographically, we consider the following AdS metric in Poincar\'e patch
\begin{equation}
    \dd s_{\mathrm{AdS}}^2 = \frac{L^2}{z^2}\left(\dd z^2 + \dd x^i\dd x^i\right) = \frac{L^2}{z^2}\left(\dd z^2 - \dd x^+\dd x^- + \dd x^{\bar i}\dd x^{\bar i}\right).
\end{equation}
Inspired by \eqref{eq:bdr-y-to-x}, we define the new bulk coordinates as follows
\begin{equation}
    u = L\frac{z}{x^+}, \qquad y^+ = -\frac{L^2}{x^+}, \qquad y^- = x^- - \frac{z^2}{x^+} - \frac{x^{\bar i}x^{\bar i}}{x^+}, \qquad y^{\bar i} = L\frac{x^{\bar i}}{x^+}\,,
\end{equation}
which is an isometric transformation in the bulk
\begin{align}
    \dd s^2_{\mathrm{AdS}} & = \frac{L^2}{z^2}\left(\dd z^2 - \dd x^+\dd x^- + \dd x^{\bar i}\dd x^{\bar i}\right)= \frac{L^2}{u^2}\left(\dd u^2 - \dd y^+\dd y^- + \dd y^{\bar i}\dd y^{\bar i}\right),
\end{align}
and reproduces \eqref{eq:bdr-y-to-x} at the boundary $u = 0$. According to holographic dictionary, the energy-momentum tensor is dual to the metric perturbation $h_{ab}$. Specifically, the energy flux operator in \eqref{eq:energy-flux-op-in-y} is sourced by $\hat h_{++} = L^2\Omega^3\delta(y^+)\delta^{d - 2}(y^{\bar i} - y'^{\bar i})$, so that the bulk solution is
\begin{align}
    h_{++} & \propto \frac{L^2}{u^2}\int \dd y'^- \frac{u^d}{\left[u^2 - y^+(y^- - y'^-) + (y^{\bar i} - y'^{\bar i})(y^{\bar i} - y'^{\bar i}) + i\epsilon\right]^d}\nonumber\\
    & \propto \frac{L^2}{u^2}\frac{\delta(y^+)u^d}{\left[u^2 + (y^{\bar i} - y'^{\bar i})(y^{\bar i} - y'^{\bar i})\right]^{d - 1}}\,,\label{eq:hpp-sol}
\end{align}
where the overall factor is unimportant so we ignore it here. Remarkably, the above insertion for $\mathcal E(\vec n)$ can be done using an exact solution instead of perturbation by considering the shockwave solution
\begin{equation}\label{eq:shockwave}
    \dd s^2_{\mathrm{shockwave}} = \dd s^2_{\mathrm{AdS}} + \frac{L^2}{u^2}\delta(y^+) W(u, y^{\bar i})(\dd y^+)^2\,,
\end{equation}
where $W$ satisfies the equation of motion
\begin{equation}\label{eq:w-eom}
    \partial_u^2 W - \frac{d - 1}{u}\partial_u W + \partial_i\partial_i W = 0\,.
\end{equation}
It is important to note that this equation will not be altered by higher curvature terms \cite{Horowitz:1999gf}. We now only need to consider the second-order perturbation around the shockwave background, instead of the general third order around the AdS background. This simplifies the calculation greatly. Comparing \eqref{eq:shockwave} with \eqref{eq:hpp-sol}, the desired bulk solution of $W$ is given by
\begin{equation}\label{eq:w-sol}
    W \propto \frac{u^d}{\left[u^2 + (y^{\bar i} - y'^{\bar i})(y^{\bar i} - y'^{\bar i})\right]^{d - 1}}\,.
\end{equation}
One can verify that it indeed satisfies \eqref{eq:w-eom}.

For the excitation operator $\mathcal O$, we choose the polarization to be $\varepsilon_{x^1x^2} = \varepsilon_{x^2x^1} = 1$ with all other components vanishing, so that the only non-vanishing component of the metric perturbation is $h_{x^1x^2}$. This implies that this particular holographic procedure requires $d\geqslant 4$, even though the CFT energy flux (\ref{eq:energy-flux-general-form}) can be defined in $d=3$. Since $h_{++}$ is localized at $y^+ = 0$, we are only interested in the behavior of $h_{x^1x^2}$ on this surface. It can be shown \cite{Hofman:2008ar, Myers:2010jv} that after transforming to $(u, y^a)$ coordinate, $h_{y^1y^2}$ is also localized, namely
\begin{equation}\label{eq:h12-behaviour}
    h_{y^1y^2}(u, y^+ = 0, y^-, y^{\bar i}) \sim e^{-iEy^-/2}\delta(u - L)\delta^{d - 2}(y^{\bar i})\,.
\end{equation}
The transformation also intruduces other components $h_{y^+y^1}$, $h_{y^+y^2}$, $h_{y^+y^+}$, but as we shall see later, they can be eliminated by imposing the transverse and traceless condition
\begin{equation}
    h_a^a = 0\,, \qquad \nabla_a h^{ab} = 0\,.
\end{equation}
Defining $\phi$ by $h_{y^1y^2} = (L^2/u^2)\phi$ and imposing the transverse and traceless condition, the equation of motion of $\phi$ is
\begin{equation}\label{eq:phi-eom}
    \partial_u^2\phi - \frac{d - 1}{u}\partial_u\phi - 4\partial_{y^+}\partial_{y^-}\phi + \partial_{\bar i}\partial_{\bar i}\phi = 0\,,
\end{equation}
up to interaction terms with the shockwave.

With these preliminaries, we are ready to evaluate the energy flux. This can be done by turning on the perturbations on the shockwave metric and evaluate the on-shell action, and then extract the terms of the form $W\phi^2$. Note that by \eqref{eq:h12-behaviour} the bulk coordinate $u$ is localized at $u = L$, so we do not need to consider the boundary action. After imposing the transverse and traceless condition, using the equations of motion \eqref{eq:w-eom}, \eqref{eq:phi-eom}, and integration by parts, the on-shell effective action becomes
\begin{equation}\label{eq:cubic-s-eff}
    S^{(3)} = -\frac{1}{L^{d + 1}}\int\dd^{d + 1}x\phi\partial_{y^-}^2\phi W\left(\hat C_T + \hat t_2 T_2 + \hat t_4 T_4\middle)\right|_{u = L, y^{\bar i} = 0}\,.
\end{equation}
The basis functions $T_2$ and $T_4$ depend only on the shockwave metric function $W$, namely
\begin{align}
    T_2 & = \frac{1}{2d(d - 1)W}\big[u^2(\partial_{y^1}^2 W + \partial_{y^2}^2 W) - 2u\partial_u W\big],\\
    T_4 & = \frac{2}{d(d - 1)(d - 2)(d + 1)W}\Big[(d - 1)u^2(\partial_{y^1}^2W + \partial_{y^2}^2W) + u^4\partial_{y^1}^2\partial_{y^2}^2W - (d + 2)u\partial_u W\nonumber\\
    & \qquad - u^3(\partial_u\partial_{y^1}^2 W + \partial_u\partial_{y^2}^2W) - u^2\partial_{\hat i}\partial_{\hat i}W\Big],
\end{align}
where the index $\hat i$ covers the remaining $(d - 4)$ directions, {\it i.e.}, $3 \leqslant \hat i \leqslant d - 1$. Substitute the solution \eqref{eq:w-sol} of $W$ into the above leads to
\begin{equation}
    T_2 = \frac{n_1^2 + n_2^2}{2} - \frac{1}{d - 1}, \qquad T_4 = 2n_1^2n_2^2 - \frac{2}{d^2 - 1}.
\end{equation}
While $T_2$ and $T_4$ are independent of the detail of the action, the coefficients $\hat C_T$, $\hat t_2$, $\hat t_4$ are determined by the coupling constants of higher curvature gravities. Specifically, we find
\begingroup\allowdisplaybreaks
\begin{align}
    \hat C_T & = L^{d - 1}\Big\{1 + 2L^{-2}\big[-d (d+1) e_{2,1}-d e_{2,2}+2 (d-3) e_{2,3}\big] + L^{-4}\big[3 (d+1)^2 d^2 e_{3,1}\nonumber\\
    & \qquad +3 (d+1) d^2 e_{3,2}+3 d^2 e_{3,3}+3 d^2 e_{3,4}-2 (d+1) (2 d-7) d e_{3,5}-2 (2 d-7) d e_{3,6}\nonumber\\
    & \qquad +(60-24 d) e_{3,7}+3 (3 d-5) e_{3,8}\big] + L^{-6}\big[-2 d^3 (d+1)^3 e_{4,1}-2 d^3 (d+1)^2 e_{4,2}\nonumber\\
    & \qquad -2 d^3 (d+1) e_{4,3}-2 d^3 (d+1) e_{4,4}-2 d^3 e_{4,5}-2 d^3 (d+1) e_{4,6}-2 d^3 e_{4,7}\nonumber\\
    & \qquad +2 (d-4) d^2 (d+1)^2 e_{4,8}+2 (d-4) d^2 (d+1) e_{4,9}+2 (d-4) d^2 (d+1) e_{4,10}\nonumber\\
    & \qquad +2 (d-4) d^2 e_{4,11}+2 (d-4) d^2 e_{4,12}-2 d^3 e_{4,13}+2 (d-4) d^2 e_{4,14}\nonumber\\
    & \qquad +4 d (d+1) (3 d-8) e_{4,15}-d (d+1) (5 d-8) e_{4,16}+2 (d-4) d^2 e_{4,17}+4 d (3 d-8) e_{4,18}\nonumber\\
    & \qquad -d (5 d-8) e_{4,19}+8 (d-3) d (d+1) e_{4,20}+8 (d-3) d e_{4,21}+16 (3 d-7) e_{4,22}\nonumber\\
    & \qquad +8 (3 d-7) e_{4,23}+2 \left(d^2-10 d+14\right) e_{4,24}-2 \left(d^2-7 d+14\right) e_{4,25}\nonumber\\
    & \qquad +\frac{1}{2} \left(8 d^2-52 d+56\right) e_{4,26}\big]\Big\},\label{eq:hat-ct-value}
\end{align}
\begin{align}
    \hat t_2 & = d(d - 1)L^{d - 1}\Big\{4 L^{-2} e_{2,3} + L^{-4}[-4 d (d+1) e_{3,5}-4 d e_{3,6}-36 (d+2) e_{3,7}-3 (7 d+2) e_{3,8}]\nonumber\\
    & \qquad L^{-6}[4 (d+1)^2 d^2 e_{4,8}+4 (d+1) d^2 e_{4,9}+4 (d+1) d^2 e_{4,10}+4 d^2 e_{4,11}+4 d^2 e_{4,12}+4 d^2 e_{4,14}\nonumber\\
    & \qquad +4 d^2 e_{4,17}+36 (d+1) (d+2) d e_{4,15}+3 (d+1) (7 d+2) d e_{4,16}+36 (d+2) d e_{4,18}\nonumber\\
    & \qquad +3 (7 d+2) d e_{4,19}+16 (d+1) d e_{4,20}+16 d e_{4,21}+96 (3 d+5) e_{4,22}+48 (3 d+5) e_{4,23}\nonumber\\
    & \qquad +4 (13 d-6) e_{4,24}+(24-48 d) e_{4,25}+8 (7 d-3) e_{4,26}]\Big\},\label{eq:hat-t2-value}
\end{align}
\begin{align}
    \hat t_4 & = 6 d (d+1) (d-1) (d+2) L^{d - 1}\Big\{L^{-4}[2 e_{3,7}+e_{3,8}] + L^{-6}[-2 d (d+1) e_{4,15}-d (d+1) e_{4,16}\nonumber\\
    & \qquad -2 d e_{4,18}-d e_{4,19}-16 e_{4,22}-8 e_{4,23}-2 e_{4,24}+2 e_{4,25}-2 e_{4,26}]\Big\}.\label{eq:hat-t4-value}
\end{align}\endgroup
To obtain the final result, we need to divide the cubic action by the two-point function $\braket{T_{12}T_{12}}$, which is proportional to $C_T$. The latter can be calculated using \eqref{eq:ct-from-kappaeff} and for the case we study, we find
\begin{equation}
    C_T = \frac{2\Gamma(d + 2)}{\pi^{d/2}(d - 1)\Gamma(d / 2)}\hat C_T\,.
\end{equation}
Thus we have
\begin{equation}\label{eq:cubic-s-divide-ct}
    \frac{S^{(3)}}{\braket{T_{12}T_{12}}} \sim 1 + \frac{\hat t_2}{\hat C_T}\left(\frac{n_1^2 + n_2^2}{2} - \frac{1}{d - 1}\right) + \frac{\hat t_4}{\hat C_T}\left(2n_1^2n_2^2 - \frac{2}{d^2 - 1}\right).
\end{equation}
On the other hand, specializing to the polarization $\varepsilon_{x^1x^2} = \varepsilon_{x^2x^1} = 1$, the two terms involving $\vec n$ in \eqref{eq:energy-flux-general-form} become
\begin{equation}
    \frac{\varepsilon_{ij}^\ast\varepsilon_{il}n^jn^l}{\varepsilon_{ij}^\ast\varepsilon_{ij}} = \frac{n_1^2 + n_2^2}{2}\,, \qquad \frac{\left|\varepsilon_{ij}n^in^j\right|^2}{\varepsilon_{ij}^\ast\varepsilon_{ij}} = 2n_1^2 n_2^2\,.
\end{equation}
By comparing \eqref{eq:cubic-s-divide-ct} to \eqref{eq:energy-flux-general-form}, we arrive at the final result
\begin{equation}\label{eq:t2-t4-result}
    t_2 = \frac{\hat t_2}{\hat C_T}\,, \qquad t_4 = \frac{\hat t_4}{\hat C_T}\,,
\end{equation}
where the hatted variables are given by (\ref{eq:hat-ct-value}), (\ref{eq:hat-t2-value}) and (\ref{eq:hat-t4-value}).

Now we examine some known special cases. Firstly, for Lovelock gravities up to and including the quartic order, we set the coupling constants to
\begingroup\allowdisplaybreaks
\begin{align}
    \{e_{2, i}\} & = \mu_2\left[\prod_{i = 2}^3(d - i)\right]^{-1}\{1, -4, 1\}\,,\\
    \{e_{3, i}\} & = \mu_3\left[\prod_{i = 2}^5(d - i)\right]^{-1}\{1,-12,16,24,3,-24,4,-8\}\,,\\
    \{e_{4, i}\} & = \mu_4\left[\prod_{i = 2}^7(d - i)\right]^{-1}\{1,-24,64,48,-96,96,-384,6,-96,-24,192,96,-192,192,\nonumber\\
    & \qquad 16,-32,192,-192,384,3,-48,6,48,-96,48,-96\}\,.
\end{align}\endgroup
Substituting them into our results \eqref{eq:t2-t4-result}, \eqref{eq:hat-ct-value}, \eqref{eq:hat-t2-value}, \eqref{eq:hat-t4-value}, we obtain
\begin{equation}\label{eq:lovelock-t2-t4-ours}
    t_2 = \frac{4 (d-1) d}{(d-3) (d-2)} \frac{\mu_2 L^{-2}-3 \mu_3 L^{-4} + 6 \mu_4L^{-6}}{1-2 \mu_2 L^{-2}+3 \mu_3 L^{-4}-4 \mu_4L^{-6}}, \qquad t_4 = 0.
\end{equation}
Energy flux parameters for general Lovelock gravity was derived in ref.\cite{Camanho:2013pda}. Explicitly we have (after adapting to our conventions)
\begin{equation}\label{eq:lovelock-t2-t4}
    t_2 = \frac{2d(d - 1)}{(d - 2)(d - 3)}\frac{h''(-L^{-2})}{L^2h'(-L^{-2})}\,, \qquad t_4 = 0\,.
\end{equation}
In our case the function $h(\lambda)$ is given by
\begin{equation}
    h(\lambda) = d(d - 1)\left(\frac{1}{L_0^2} + \lambda + \mu_2\lambda^2 + \mu_3\lambda^3 + \mu_4\lambda^4\right).
\end{equation}
Substituting this into \eqref{eq:lovelock-t2-t4} we get exactly identical result with \eqref{eq:lovelock-t2-t4-ours}.

Secondly, after specializing our result to general massless cubic curvature gravity in $d = 4$ and eliminating $e_{3, 7}$, $e_{3, 8}$ by the massless condition (\ref{cubicmassless}), we arrive at
\begin{align}
    t_2 & = \frac{48 \left(2340 e_{3,1}+552 e_{3,2}+144 e_{3,3}+123 e_{3,4}+316 e_{3,5}+80 e_{3,6}\right)}{L^4-2 \left(60 e_{3,1}+8 e_{3,2}+e_{3,4}+4 e_{3,5}\right)},\\
    t_4 & = -\frac{360 \left(600 e_{3,1}+140 e_{3,2}+36 e_{3,3}+31 e_{3,4}+80 e_{3,5}+20 e_{3,6}\right)}{L^4-2 \left(60 e_{3,1}+8 e_{3,2}+e_{3,4}+4 e_{3,5}\right)},
\end{align}
which reproduces the result of ref.\cite{Li:2019auk}.

We can now obtain all the three-point function parameters ($\mathcal A$, $\mathcal B$, $\mathcal C$) by inverting \eqref{eq:ct-from-abc}, \eqref{eq:t2-from-abc}, \eqref{eq:t4-from-abc}. The final expressions of ($\mathcal A$, $\mathcal B$, $\mathcal C)$ in general dimension $d$ are recorded in appendix \ref{sec:ap:abc}.

\section{Discussions}
\label{sec:discussions}

Having obtained all the three-point function parameters ${\cal (A, B, C)}$ holographically from higher curvature gravities up to and including the quartic order, we can compare our results to those in literature and study their implications.

Firstly it is interesting to mention that even though our result, based on its derivation, should be valid only for $d \geqslant 4$, there exists a smooth $d=3$ limit. In particular, if
we restrict to massless cubic gravity and set $d = 3$, our result gives rise to the following value for $t_4$
\begin{equation}
    t_4 = -\frac{120 \left(360 e_{3,1}+96 e_{3,2}+27 e_{3,3}+25 e_{3,4}+64 e_{3,5}+18 e_{3,6}\right)}{L^4-2 \left(36 e_{3,1}+6 e_{3,2}+e_{3,4}+4 e_{3,5}\right)}
\end{equation}
which is precisely the one obtained in ref.\cite{Li:2019auk}that was derived using a different method. We thus expect that our new result of $t_4$, arising from the quartic massless gravity, is also valid at $d=3$. From our general results, setting $d = 3$ also gives a nonzero value for $t_2$; however, in $d = 3$ the symmetry group reduces to $\mathrm O(2)$, and hence there can only be one energy flux parameter $t_4$. Furthermore, specializing $\hat t_2$ in $d = 3$, we find it is actually linearly independent of $\hat C_T$ and $\hat t_4$. Therefore, it is interesting to explore the physical meaning of the value of $t_2$ of general $d$ in the $d = 3$ limit.

Secondly, we examine the conjecture \eqref{eq:t2-t4-hpp-relation}. Since for massless gravity $h'(-L^{-2})$ is proportional to $C_T$ \cite{Bueno:2018yzo}, it is equivalent to that $\hat t_2$, $\hat t_4$ and $h''(-L^{-2})$ are linearly dependent, {\it i.e.},
\begin{equation}
    a_{(d)}'\hat t_2 + b_{(d)}'\hat t_4 = h''(-L^{-2})\,.
\end{equation}
However, we find that they are actually linearly independent for general massless higher curvature gravities. As mentioned earlier, the conjecture \eqref{eq:t2-t4-hpp-relation} already was verified for certain special cases, it is thus natural to expect that there exists a more general case that satisfies this conjecture. Although the most general and unified condition for all orders that satisfies \eqref{eq:t2-t4-hpp-relation} is hard to come by, we actually find that \eqref{eq:t2-t4-hpp-relation} is satisfied by cubic and quartic \ac{qt} gravities in general dimension, with coefficients given by
\begin{equation}
    a_{(d)}  = -\frac{(d-3) (d-2)}{2 (d-1) d},\qquad b_{(d)}  = -\frac{(d-2) \left(7 d^3-19 d^2-8 d+8\right)}{2 (d-1) d (d+1) (d+2) (2 d-1)}.
\end{equation}
The values of $a_{(d)}$ and $b_{(4)} = -1/21$ are in consistency with ref.\cite{Bueno:2018yzo} obtained from Lovelock and $d = 4$ \ac{qt} gravities. We can then reasonably believe that all \ac{qt} gravities satisfies the conjecture \eqref{eq:t2-t4-hpp-relation}. A brief review on \Ac{qt} gravities can be found in appendix \ref{sec:ap:qtg}, where we find that there are 15 such theories in quartic gravities.

Thirdly, we focus on identities involving the $a$-charge \eqref{eq:a-charge-to-abc} and \eqref{eq:ct-a-hidden-rel}. As mentioned earlier, the $a$-charge can be generalized to arbitrary dimensions as the entanglement entropy across a spherical region $S^{d - 2}$. It was shown with a conformal map that the entanglement entropy over the spherical region in Minkowski background equals to the thermal entropy of $R \times H^{d - 1}$ background. The latter can be calculated holographically by the black hole entropy of a locally AdS hyperbolic topological black hole \cite{Casini:2011kv}
\begin{equation}\label{eq:topo-bh}
    \dd s^2 = -f(r)\dd t^2 + \frac{1}{f(r)}\dd r^2 + r^2\dd\Sigma_{d - 1, k = -1}^2\,, \qquad f(r) = \frac{r^2}{L^2} - 1\,.
\end{equation}
The black hole entropy can be calculated from the Wald entropy in a standard way \cite{Wald:1993nt, Iyer:1994ys}
\begin{equation}
    S_{\mathrm{Wald}} = -2\pi\int\dd^{d - 1}x\left.\sqrt{\sigma}P^{abcd}\epsilon_{ab}\epsilon_{cd}\right|_{r = L}\,,
\end{equation}
where $\sigma_{ab}$ is the induced metric of the horizon, and $\epsilon = \dd t\wedge\dd r$ is the binormal of the horizon, satisfying $\epsilon_{ab}\epsilon^{ab} = -2$. For the topological black hole \eqref{eq:topo-bh}, the integrand is a constant proportional to the area of the horizon so the value diverges, thus one may assign the entropy density on the horizon to the $a$-charge as follows \cite{Li:2021jfh}
\begin{equation}\label{eq:a-charge-from-ee}
    a = \frac{\pi^{d / 2}}{2\pi\Gamma(d / 2)}\frac{S_{\mathrm{Wald}}}{\Omega_{d - 1, -1}} = -\frac{\pi^{d / 2}}{\Gamma(d / 2)}L^{d - 1}\left(P^{abcd}\epsilon_{ab}\epsilon_{cd}\right)_{r = L}.
\end{equation}
It can be shown that with this definition the value coincides with $a$-charge in even dimensions. It is straightforward to calculate the $a$-charge from \eqref{eq:a-charge-from-ee}, we obtain
\begingroup\allowdisplaybreaks
\begin{align}
    a & = \frac{2\pi^{d / 2}}{\Gamma(d / 2)} L^{d - 1}\Big\{1 - 2L^{-2}\big[d (d+1) e_{2,1} + d e_{2,2} + 2 e_{2,3}\big] + 3L^{-4}\big[d^2 (d+1)^2 e_{3,1}\nonumber\\
    & \qquad + d^2 (d+1) e_{3,2} + d^2 e_{3,3} + d^2 e_{3,4} + 2 d (d+1) e_{3,5} + 2 d e_{3,6} + 4 e_{3,7} + (d-1) e_{3,8}\big]\nonumber\\
    & \qquad + 4L^{-6}\big[-d^3 (d+1)^3 e_{4,1}-d^3 (d+1)^2 e_{4,2}-d^3 (d+1) e_{4,3}-d^3 (d+1) e_{4,4}-d^3 e_{4,5}\nonumber\\
    & \qquad -d^3 (d+1) e_{4,6}-d^3 e_{4,7}-2 d^2 (d+1)^2 e_{4,8}-2 d^2 (d+1) e_{4,9}-2 d^2 (d+1) e_{4,10}-2 d^2 e_{4,11}\nonumber\\
    & \qquad -2 d^2 e_{4,12}-d^3 e_{4,13}-2 d^2 e_{4,14}-4 d (d+1) e_{4,15}-d \left(d^2-1\right) e_{4,16}-2 d^2 e_{4,17}-4 d e_{4,18}\nonumber\\
    & \qquad -(d-1) d e_{4,19}-4 d (d+1) e_{4,20}-4 d e_{4,21}-8 e_{4,22}-4 e_{4,23}-2 (d-1) e_{4,24}\nonumber\\
    & \qquad -\left(d^2-d+2\right) e_{4,25}+(2-3 d) e_{4,26}\big]\Big\}.\label{eq:a-charge-value}
\end{align}\endgroup
With both $C_T$ and $a$ evaluated, it follows immediately that after applying the massless condition we have \eqref{eq:ct-a-hidden-rel}.

Specializing our results to $d = 4$, we reproduce \eqref{eq:a-charge-to-abc} and therefore we verify the relation holographically in higher curvature gravity up to and including the quartic order. However, for $d \geqslant 5$, we find that $a$ and ($\mathcal A$, $\mathcal B$, $\mathcal C$) are in general  linearly independent. In other words, the $d=4$ relation \eqref{eq:a-charge-to-abc} does not have a higher-dimensional generalization. This somewhat unexpected result instructs us to further consider central charges in $d = 6$, where there are three $c$-charges. Details and explicit values of the $c$-charges in $d = 6$ can be found in appendix \ref{sec:ap:central-charges-6d}. We find that all three $c$-charges turn out be linear combinations of ($\mathcal A$, $\mathcal B$, $\mathcal C$), namely
\begin{align}
    c_1 & = \frac{\pi ^9}{233280} (-98 \mathcal{A}+9 \mathcal{B}+174 \mathcal{C})\,, \\
    c_2 & = \frac{\pi ^9}{6531840} (1226 \mathcal{A}-153 \mathcal{B}+882 \mathcal{C})\,, \\
    c_3 & = \frac{\pi^6}{3024}\left.C_T\right|_{d = 6} = \frac{\pi ^9}{145152} (20 \mathcal{A}-\mathcal{B}-14 \mathcal{C})\,.
\end{align}
In terms of $t_2$, $t_4$, we have
\begin{equation}
    t_2  = \frac{15 \left(23 c_1-44 c_2+144 c_3\right)}{16 c_3},\qquad
    t_4  = -\frac{105 \left(c_1-2 c_2+6 c_3\right)}{2 c_3}\,,
\end{equation}
which coincides with the CFT results derived from free Dirac fermion, real scalar, and antisymmetric two-form fields \cite{Osborn:2015rna}. Our results suggest that this is indeed a universal property of $d=6$ CFT.

\section{Conclusion}
\label{sec:conclusion}

In this work we considered general higher curvature gravity up to and including the quartic order and calculated holographically the three-point function parameters ($\mathcal A$, $\mathcal B$, $\mathcal C$) of the dual field theory in general $d$ dimensions. We adopted an indirect method by calculating the central charge $C_T$ and the energy flux parameters $t_2$, $t_4$, which are known to be directly related to the three-point function parameters. We therefore obtained the complete list of the holographic results of $(a, C_T)$ and $(\cal A, B, C)$ for general $d$ dimensions.

Despite of the fact that our method should be valid only for $d\geqslant 4$, we not only reproduce the previously known result in $d=4$, but also the correct value of $t_4$ of ref.\cite{Li:2019auk} after setting $d=3$. Therefore it may be of interest to explore the physical meaning of $t_2$ in this case since at $d = 3$ the parameter $t_2$ does not contribute to the energy flux.

We also examined the relation between $t_2$, $t_4$, $h''$ proposed by ref.\cite{Bueno:2018yzo} and found that it does not hold for general massless gravities, but it is satisfied by \ac{qt} gravity. We obtained the missing coefficient $b_{(d)}$ for general $d$, extending the previously known $d=4$ result. We also found that the $d=4$ identity (\ref{eq:a-charge-to-abc}) cannot be generalized to higher dimensions. The generalized $a$-charge is linearly independent of $(\cal A, B, C)$ when $d \geqslant 5$.  We calculated the $c$-charges in $d = 6$ and found they are all linear combinations of $(\cal A, B, C)$. Considering the fact that the number of $c$-charges proliferates as $d$ increases, their relations to $(\cal A, B, C)$ are hard to conjecture and require a new investigation.

\section{Acknowledgement}

The work is supported in part by the National Natural Science Foundation of China (NSFC) grants No.~11935009 and No.~12375052.

\appendix

\section{Massless condition for quartic curvature gravity}
\label{sec:ap:quartic-massless-cond}

In this appendix, we give the conditions on the coupling constants for higher curvature massless gravity. The condition for the decoupling of the massive scalar mode is
\begin{align}
    & (d+1)^2 d^2 e_{4,2}+3 (d+1) d^2 e_{4,3}+2 (d+1) d^2 e_{4,4}+6 d^2 e_{4,5}+4 (d+1)^2 d^2 e_{4,8}\nonumber\\
    & +\left(3 d^2-2 d+1\right) e_{4,13}+\left(4 d^2+9 d-1\right) e_{4,14}+2 \left(2 d^2+3 d-2\right) e_{4,17}\nonumber\\
    & +(d+1) (2 d-1) d e_{4,6}+2 (2 d-1) d e_{4,7}+4 (d+1)^2 d e_{4,9}+2 (d+1) (2 d+1) d e_{4,10}\nonumber\\
    & +2 (2 d+5) d e_{4,11}+24 (d+1) d e_{4,15}-3 (d+1) d e_{4,16}+16 (d+1) d e_{4,20}+\left(4 (d+1)^2-6\right) e_{4,12}\nonumber\\
    & +12 (2 d+1) e_{4,18}-(d+4) e_{4,19}+16 (d+1) e_{4,21}+2 (2 d-7) e_{4,24}+4 (d+4) e_{4,25}\nonumber\\
    & +2 (4 d-5) e_{4,26}+96 e_{4,22}+48 e_{4,23} = 0.
\end{align}
The condition for the decoupling of the massive ghostlike spin-2 mode is
\begin{align}
    & 24 (d+1)^2 d^2 e_{4,1}+(d+1) (d+21) d^2 e_{4,2}+3 (d+5) d^2 e_{4,3}+2 (d+9) d^2 e_{4,4}+6 d^2 e_{4,5}\nonumber\\
    & +\left(2 d^2+17 d+3\right) d e_{4,6}+\left(3 d^2+10 d-3\right) e_{4,13}+3 \left(d^2+5 d-4\right) e_{4,16}+2 (2 d+3) d e_{4,7}\nonumber\\
    & +40 (d+1) d e_{4,8}+4 (d+7) d e_{4,9}+2 (d+17) d e_{4,10}+10 d e_{4,11}+5 d e_{4,19}+4 d e_{4,25}\nonumber\\
    & +2 (4 d+3) e_{4,12}+3 (3 d+1) e_{4,14}+6 (d+2) e_{4,17}+48 e_{4,15}+12 e_{4,18}+64 e_{4,20}+16 e_{4,21}\nonumber\\
    & +10 e_{4,24}+14 e_{4,26} = 0.
\end{align}

\section{Explicit expressions for ($\mathcal A$, $\mathcal B$, $\mathcal C$)}
\label{sec:ap:abc}
In section \ref{sec:efparams-calc}, we obtained the energy flux parameters $t_2$ and $t_4$ and also $C_T$. From these, we can read off the three-point function parameters. They are given by
\begingroup\allowdisplaybreaks
\begin{align}
    \mathcal A & = \frac{2d^3 \pi ^{-d} L^{d-1} \Gamma (d+2)}{(d-1)^3 (d+1)}\Big\{-1 + 2L^{-2}\big[d (d+1) e_{2,1}+d e_{2,2}+8 e_{2,3}\big] - 3L^{-4}\big[d^2 (d+1)^2 e_{3,1}\nonumber\\
    & \qquad + d^2 (d+1) e_{3,2} + d^2 e_{3,3} + d^2 e_{3,4} + 6 d (d+1) e_{3,5} + 6 d e_{3,6} + 36 e_{3,7} + \left(d^2-2 d-7\right) e_{3,8}\big]\nonumber\\
    & \qquad + L^{-6}\big[4 d^3 (d+1)^3 e_{4,1}+4 d^3 (d+1)^2 e_{4,2}+4 d^3 (d+1) e_{4,3}+4 d^3 (d+1) e_{4,4}+4 d^3 e_{4,5}\nonumber\\
    & \qquad +4 d^3 (d+1) e_{4,6}+4 d^3 e_{4,7}+20 d^2 (d+1)^2 e_{4,8}+20 d^2 (d+1) e_{4,9}+20 d^2 (d+1) e_{4,10}\nonumber\\
    & \qquad +20 d^2 e_{4,11}+20 d^2 e_{4,12}+4 d^3 e_{4,13}+20 d^2 e_{4,14}+112 d (d+1) e_{4,15}\nonumber\\
    & \qquad +d (d+1) \left(3 d^2-5 d-22\right) e_{4,16}+20 d^2 e_{4,17}+112 d e_{4,18}+d \left(3 d^2-5 d-22\right) e_{4,19}\nonumber\\
    & \qquad +64 d (d+1) e_{4,20}+64 d e_{4,21}+512 e_{4,22}+256 e_{4,23}+4 \left(3 d^2-4 d-26\right) e_{4,24}\nonumber\\
    & \qquad -8 \left(d^2-4 d-13\right) e_{4,25}+4 \left(3 d^2-26\right) e_{4,26}\big]\Big\}.
\end{align}
\begin{align}
    \mathcal B & = \frac{d \pi ^{-d} L^{d-1} \Gamma (d+2)}{(d-1)^3 (d+1)}\Big\{-2 \left(d^3-d^2+1\right) + 4L^{-2}\big[d (d+1) \left(d^3-d^2+1\right) e_{2,1}\nonumber\\
    & \qquad + d \left(d^3-d^2+1\right) e_{2,2}-\left(d^4-7 d^3+5 d^2-d-6\right) e_{2,3}\big]\nonumber\\
    & \qquad + L^{-4}\big[-6 d^2 (d+1)^2 \left(d^3-d^2+1\right) e_{3,1}-6 d^2 (d+1) \left(d^3-d^2+1\right) e_{3,2}\nonumber\\
    &\qquad -6 d^2 \left(d^3-d^2+1\right) e_{3,3}-6 d^2 \left(d^3-d^2+1\right) e_{3,4}+4 d (d+1) \left(d^4-8 d^3+6 d^2-d-7\right) e_{3,5}\nonumber\\
    & \qquad +4 d \left(d^4-8 d^3+6 d^2-d-7\right) e_{3,6}-12 \left(d^5-3 d^4+11 d^3-7 d^2+6 d+10\right) e_{3,7}\nonumber\\
    & \qquad -3 \left(3 d^5-3 d^4-11 d^3+9 d^2-4 d-10\right) e_{3,8}\big] + L^{-6}\big[8 d^3 (d+1)^3 \left(d^3-d^2+1\right) e_{4,1}\nonumber\\
    & \qquad +8 d^3 (d+1)^2 \left(d^3-d^2+1\right) e_{4,2}+8 d^3 (d+1) \left(d^3-d^2+1\right) e_{4,3}\nonumber\\
    & \qquad +8 d^3 (d+1) \left(d^3-d^2+1\right) e_{4,4}+8 d^3 \left(d^3-d^2+1\right) e_{4,5}+8 d^3 (d+1) \left(d^3-d^2+1\right) e_{4,6}\nonumber\\
    & \qquad +8 d^3 \left(d^3-d^2+1\right) e_{4,7}-4 d^2 (d+1)^2 \left(d^4-9 d^3+7 d^2-d-8\right) e_{4,8}\nonumber\\
    & \qquad -4 d^2 (d+1) \left(d^4-9 d^3+7 d^2-d-8\right) e_{4,9}-4 d^2 (d+1) \left(d^4-9 d^3+7 d^2-d-8\right) e_{4,10}\nonumber\\
    & \qquad -4 d^2 \left(d^4-9 d^3+7 d^2-d-8\right) e_{4,11}-4 d^2 \left(d^4-9 d^3+7 d^2-d-8\right) e_{4,12}\nonumber\\
    & \qquad +8 d^3 \left(d^3-d^2+1\right) e_{4,13}-4 d^2 \left(d^4-9 d^3+7 d^2-d-8\right) e_{4,14}\nonumber\\
    & \qquad +4 d (d+1) \left(3 d^5-9 d^4+35 d^3-23 d^2+18 d+32\right) e_{4,15}\nonumber\\
    & \qquad +d (d+1) (d+2) \left(9 d^4-25 d^3+13 d^2+3 d-16\right) e_{4,16}\nonumber\\
    & \qquad -4 d^2 \left(d^4-9 d^3+7 d^2-d-8\right) e_{4,17}\nonumber\\
    & \qquad +4 d \left(3 d^5-9 d^4+35 d^3-23 d^2+18 d+32\right) e_{4,18}\nonumber\\
    & \qquad +d (d+2) \left(9 d^4-25 d^3+13 d^2+3 d-16\right) e_{4,19}\nonumber\\
    & \qquad -16 d (d+1) \left(d^4-7 d^3+5 d^2-d-6\right) e_{4,20}-16 d \left(d^4-7 d^3+5 d^2-d-6\right) e_{4,21}\nonumber\\
    & \qquad +32 \left(3 d^5-6 d^4+14 d^3-8 d^2+15 d+14\right) e_{4,22}\nonumber\\
    & \qquad +16 \left(3 d^5-6 d^4+14 d^3-8 d^2+15 d+14\right) e_{4,23}\nonumber\\
    & \qquad +4 \left(5 d^5-3 d^4-29 d^3+23 d^2-22 d-28\right) e_{4,24}\nonumber\\
    &\qquad -16 \left(d^5-2 d^4-6 d^3+5 d^2-7 d-7\right) e_{4,25}\nonumber\\
    & \qquad +8 \left(2 d^5+2 d^4-17 d^3+12 d^2-8 d-14\right) e_{4,26}\big]\Big\}.
\end{align}
\begin{align}
    \mathcal C & = \frac{d^2 \pi ^{-d} L^{d-1} \Gamma (d+2)}{2(d-1)^3 (d+1)}\Big\{-2 d^2+2 d+1 + 2L^{-2}\big[d (d+1) \left(2 d^2-2 d-1\right) e_{2,1}\nonumber\\
    & \qquad +d \left(2 d^2-2 d-1\right) e_{2,2}-2 \left(d^3-8 d^2+5 d+6\right) e_{2,3}\big]\nonumber\\
    & \qquad + L^{-4}\big[-3 d^2 (d+1)^2 \left(2 d^2-2 d-1\right) e_{3,1}-3 d^2 (d+1) \left(2 d^2-2 d-1\right) e_{3,2}\nonumber\\
    & \qquad -3 d^2 \left(2 d^2-2 d-1\right) e_{3,3}-3 d^2 \left(2 d^2-2 d-1\right) e_{3,4}\nonumber\\
    & \qquad +2 d (d+1) \left(2 d^3-18 d^2+12 d+13\right) e_{3,5}+2 d \left(2 d^3-18 d^2+12 d+13\right) e_{3,6}\nonumber\\
    & \qquad +36 \left(d^3-6 d^2+3 d+5\right) e_{3,7}-3 \left(d^4-2 d^3-8 d^2+2 d+15\right) e_{3,8}\big]\nonumber\\
    & \qquad + L^{-6}\big[4 d^3 (d+1)^3 \left(2 d^2-2 d-1\right) e_{4,1}+4 d^3 (d+1)^2 \left(2 d^2-2 d-1\right) e_{4,2}\nonumber\\
    & \qquad +4 d^3 (d+1) \left(2 d^2-2 d-1\right) e_{4,3}+4 d^3 (d+1) \left(2 d^2-2 d-1\right) e_{4,4}+4 d^3 \left(2 d^2-2 d-1\right) e_{4,5}\nonumber\\
    & \qquad +4 d^3 (d+1) \left(2 d^2-2 d-1\right) e_{4,6}+4 d^3 \left(2 d^2-2 d-1\right) e_{4,7}\nonumber\\
    & \qquad -4 d^2 (d+1)^2 \left(d^3-10 d^2+7 d+7\right) e_{4,8}-4 d^2 (d+1) \left(d^3-10 d^2+7 d+7\right) e_{4,9}\nonumber\\
    & \qquad -4 d^2 (d+1) \left(d^3-10 d^2+7 d+7\right) e_{4,10}-4 d^2 \left(d^3-10 d^2+7 d+7\right) e_{4,11}\nonumber\\
    & \qquad -4 d^2 \left(d^3-10 d^2+7 d+7\right) e_{4,12}\nonumber\\
    & \qquad +4 d^3 \left(2 d^2-2 d-1\right) e_{4,13}-4 d^2 \left(d^3-10 d^2+7 d+7\right) e_{4,14}\nonumber\\
    & \qquad -4 d (d+1) \left(9 d^3-56 d^2+29 d+46\right) e_{4,15}+d (d+1) (d+2) \left(3 d^3-10 d^2-8 d+23\right) e_{4,16}\nonumber\\
    & \qquad -4 d^2 \left(d^3-10 d^2+7 d+7\right) e_{4,17}-4 d \left(9 d^3-56 d^2+29 d+46\right) e_{4,18}\nonumber\\
    & \qquad +d (d+2) \left(3 d^3-10 d^2-8 d+23\right) e_{4,19}-16 d (d+1) \left(d^3-8 d^2+5 d+6\right) e_{4,20}\nonumber\\
    & \qquad -16 d \left(d^3-8 d^2+5 d+6\right) e_{4,21}-64 \left(3 d^3-16 d^2+7 d+14\right) e_{4,22}\nonumber\\
    & \qquad -32 \left(3 d^3-16 d^2+7 d+14\right) e_{4,23}+4 \left(2 d^4+d^3-35 d^2+3 d+56\right) e_{4,24}\nonumber\\
    & \qquad -4 \left(d^4-5 d^3-29 d^2+9 d+56\right) e_{4,25}+4 \left(d^4+9 d^3-40 d^2-3 d+56\right) e_{4,26}\big]\Big\}.
\end{align}\endgroup

\section{Central $c$-charges in $d = 6$}
\label{sec:ap:central-charges-6d}

In $d = 6$ there are three Weyl invariants $I_i$ \cite{Bastianelli:2000rs}
\begin{align}
    I_1 & = C_{abcd}C^{aefd}C_e{}^{bc}{}_f, \nonumber \\
    I_2 & = C_{abcd}C^{cdef}C_{ef}{}^{ab}, \nonumber \\
    I_3 & = C_{abcd}\nabla^2C^{abcd} + 4 C_{abcd} R^a_e C^{ebcd} - \frac65 C_{abcd}C^{abcd}R + \nabla_a J^a
\end{align}
where $C_{abcd}$ is the Weyl tensor, and the explicit form of $\nabla_a J^a$ is irrelavent since it is a total divergence and can be canceled by a local counterterm. This gives three $c$-charges in $d = $6.

The central charges of cubic curvature gravity in $d = 6$ was computed in ref.\cite{Bugini:2016nvn}, we therefore extend the result to quartic order. We employ the reduced Fefferman-Graham expansion trick to calculate the central charges \cite{Henningson:1998ey, Li:2017txk}. We find that the $a$-charge is given by \eqref{eq:a-charge-value} specialized to $d = 6$, and the three $c$-charges are
\begingroup\allowdisplaybreaks
\begin{align}
    c_1 & = \frac{4}{3}\pi^3L^5\Big[-3 + 4L^{-2}(63 e_{2,1}+9 e_{2,2}-e_{2,3}) + 3L^{-4}(-5292 e_{3,1}-756 e_{3,2}-108 e_{3,3}-108 e_{3,4}\nonumber\\
    & \qquad -28 e_{3,5}-4 e_{3,6}+20 e_{3,7}-39 e_{3,8}) + 24L^{-6}(37044 e_{4,1}+5292 e_{4,2}+756 e_{4,3}+756 e_{4,4}\nonumber\\
    & \qquad +108 e_{4,5}+756 e_{4,6}+108 e_{4,7}+588 e_{4,8}+84 e_{4,9}+84 e_{4,10}+12 e_{4,11}+12 e_{4,12}+108 e_{4,13}\nonumber\\
    & \qquad +12 e_{4,14}-84 e_{4,15}+231 e_{4,16}+12 e_{4,17}-12 e_{4,18}+33 e_{4,19}-28 e_{4,20}-4 e_{4,21}-12 e_{4,22}\nonumber\\
    & \qquad -6 e_{4,23}+13 e_{4,24}+4 e_{4,25}+12 e_{4,26})\Big],
\end{align}
\begin{align}
    c_2 & = \frac{1}{3}\pi^3L^5\Big[-3 + 4L^{-2}(63 e_{2,1} + 9 e_{2,2} + 7 e_{2,3}) + 3L^{-4}(-5292 e_{3,1}-756 e_{3,2}-108 e_{3,3}-108 e_{3,4}\nonumber\\
    & \qquad -476 e_{3,5}-68 e_{3,6}+20 e_{3,7}-7 e_{3,8}) + 24L^{-6}(37044 e_{4,1}+5292 e_{4,2}+756 e_{4,3}+756 e_{4,4}\nonumber\\
    & \qquad +108 e_{4,5}+756 e_{4,6}+108 e_{4,7}+2940 e_{4,8}+420 e_{4,9}+420 e_{4,10}+60 e_{4,11}+60 e_{4,12}\nonumber\\
    & \qquad +108 e_{4,13}+60 e_{4,14}-84 e_{4,15}+63 e_{4,16}+60 e_{4,17}-12 e_{4,18}+9 e_{4,19}+196 e_{4,20}+28 e_{4,21}\nonumber\\
    & \qquad -44 e_{4,22}-22 e_{4,23}+13 e_{4,24}+12 e_{4,25}+20 e_{4,26})\Big],
\end{align}
\begin{equation}
    c_3 = \frac{\pi^6}{3024}\left.C_T\right|_{d = 6}.
\end{equation}
\endgroup

\section{General structure of three-point function}
\label{sec:ap:general-3pt}
Holographic three-point functions can be extracted from the cubic effective action of graviton on AdS background. For higher curvature gravities whose the Lagrangian depends on the metric $g_{ab}$ and $R_{abcd}$, the metric dependence becomes implicit if one chooses $R_{ab}{}^{cd}$ as the independent variable \cite{Padmanabhan:2013xyr}, {\it i.e.}, $\mathcal L = \mathcal L(R_{ab}{}^{cd})$. The general form of the cubic effective action can then be obatined by varying the action to the third order
\begin{align}
    S^{(3)} = \frac{1}{3!}\delta^3\int\dd^{d + 1}x \left.\sqrt{|g|}\mathcal L\right|_{\bar g_{ab}}
    & = \frac{1}{6}\int\dd^{d + 1}x\sqrt{|g|}\Big[P^{ab}{}_{cd}\delta^3(R_{ab}{}^{cd}) + 3C^{ab}{}_{cd}{}^{ef}{}_{gh}\delta^2(R_{ab}{}^{cd})\delta(R_{ef}{}^{gh})\nonumber\\
    & + G^{ab}{}_{cd}{}^{ef}{}_{gh}{}^{ij}{}_{kl}\delta(R_{ab}{}^{cd})\delta(R_{ef}{}^{gh})\delta(R_{ij}{}^{kl})\Big]_{\bar g_{ab}}\,,\label{eq:general-cubic-efa}
\end{align}
where $\bar g_{ab}$ is the AdS metric, tensors $P^{abcd}$ and $C^{abcdefgh}$ are given by \eqref{eq:general-eom} and \eqref{eq:lpdr3-c-def} respectively. The tensor $G^{abcdefghijkl}$ is defined by
\begin{equation}
    G^{abcdefghijkl} = \frac{\partial^3\mathcal L}{\partial R_{abcd}\partial R_{efgh}\partial R_{ijkl}}\,.
\end{equation}
Note that we have also imposed the transverse and traceless condition so that $\delta \sqrt{|g|} = (1/2)\sqrt{|g|}g^{ab}\delta g_{ab} = 0$. It now becomes clear that there are three different structures of the three-point function, controlled by the tensors $P^{abcd}$, $C^{abcdefgh}$, $G^{abcdefghijkl}$ evaluated on the AdS background respectively. The AdS background has $\bar R_{abcd} = -2/L^2 \bar g_{a[c}\bar g_{d]b}$, it follows that when evaluated on this background, these three tensors can only be built from the metric $\bar g_{ab}$, thus their forms are all rigidly fixed, with theory-dependent coefficients, {\it e.g.}, \eqref{eq:l-pd-r-form} and \eqref{eq:l-pd-r2-form}. For Einstein gravity only the contribution from $P^{abcd}$ is non-zero, while for quadratic gravity the $G^{abcdefghijkl}$ contribution vanishes. All three tensors are non-zero for cubic gravity, which can therefore enumerate all the possible structures. For the quartic or higher orders, no new structures arise; they just modify the coefficients in these three tensors.

\section{Quasi-topological gravity}
\label{sec:ap:qtg}

We shall briefly review \ac{qt} gravity and present dimension-generic cubic and quartic \ac{qt} combinations. There has been an extensive study on \ac{qt} gravity (e.g. ref.\cite{Myers:2010jv,Myers:2010ru,Dehghani:2013ldu,Cisterna:2017umf}). A \acf{qt} gravity is a type of gravity theories whose equation of motion on the special spherically symmetric metric ansatze
\begin{equation}\label{eq:sss-metric}
    \dd s^2 = -f(r)\dd t^2 + \frac{1}{f(r)}\dd r^2 + r^2\dd\Sigma_{d - 1, k}^2
\end{equation}
is algebraic in $f(r)$, {\it i.e.}, does not involve derivatives of $f(r)$. This condition is equivalent to \cite{Bueno:2019ycr}
\begin{equation}\label{eq:qt-cond}
    \left.\nabla_a P^{abcd}\right|_{f} = 0
\end{equation}
where $\dots|_f$ denotes evaluating on the metric ansatze \eqref{eq:sss-metric}. This makes the black hole solutions of \ac{qt} gravities easy to obtain, thus \ac{qt} gravity serves as a simplified model of general higher curvature gravity.

At a given curvature polynomial order, it is straightforward to derive \ac{qt} curvature combinations using constraints arise from \eqref{eq:qt-cond}. For our purpose we only consider dimension-generic case here, but it is important to note that at specific dimensions there may be more possible combinations than dimension-generic case. We have two linearly independent cubic combinations with coupling constants given by
\begin{align}
    \{e_{3, i}^{\mathcal Q, 1}\} & = \big\{d^2-2 d+3,-12 (d-1)^2,16 (d-2) d,24 \left(d^2-3 d+2\right),3 \left(d^2-2 d-1\right),\nonumber\\
    & \qquad -24 \left(d^2-3 d+1\right),4 d^2-14 d+6,0\big\},\\
    \{e_{3, i}^{\mathcal Q, 2}\} & = \big\{3 (d+1),12-36 d,48 (d-1),24 (d+1),9 d-15,-24 (d-1),0,8 \left(2 d^2-7 d+3\right)\big\}.
\end{align}
Note that Lovelock combination is a special case of \ac{qt} gravity, we thus have only one non-trivial cubic \ac{qt} combination. For the quartic order we have 15 independent combinations, the full set of them is too lengthy to be presented here so we only show first two of them.
\begin{align}
    \{e_{4, i}^{\mathcal Q, 1}\} & = \big\{1,-2 (d+1),8 (d-1),d^2-4 d+7,2-2 d^2,0,0,\cdots\big\},\\
    \{e_{4, i}^{\mathcal Q, 2}\} & = \big\{0,2 (d+1),0,-2 \left(d^2-2 d+5\right),0,-16 (d-1),8 \left(d^2-1\right),-d^2+3 d-4,\nonumber\\
    & \qquad 4 (d-1)^2,d^3-4 d^2+5 d+2,-2 (d-1)^2 (d+1),0,0,\cdots\big\}.\\
    \cdots\nonumber
\end{align}
The full set of quartic \ac{qt} combinations can be found in the supplemental material \texttt{r4QTG.wl}, which is a Wolfram Language file with further instructions included in the usage messages.

\bibliography{refs}

\end{document}